\documentclass[preprint,aps ,nofootinbib]{revtex4}
\pdfoutput=1
\usepackage[colorlinks=true,linkcolor=blue,urlcolor=blue,filecolor=black,citecolor=red,pdfstartview=FitV,pdftitle={},pdfsubject={},pdfkeywords={},pdfpagemode=None,bookmarksopen=true]{hyperref}
\usepackage{graphicx}
\usepackage{amsmath}
\usepackage{amsfonts}
\usepackage{amssymb,ulem}
\usepackage{color}%
\usepackage{slashed}
\usepackage{dcolumn}
\newcommand{\f}{\begin{equation}}
\newcommand{\ff}{\end{equation}}
\newcommand{\fa}{\begin{eqnarray}}
\newcommand{\ffa}{\end{eqnarray}}

\begin{document}
\title{Holographic fermionic spectrum with Weyl correction}
\author{Jian-Pin Wu}
\thanks{jianpinwu@mail.bnu.edu.cn}
\author{Baicheng Xu}
\thanks{xubcheng@163.com}
\author{Guoyang Fu}
\thanks{FuguoyangEDU@163.com}
\affiliation{
Department of
Physics, School of Mathematics and Physics, Bohai University, Jinzhou 121013, China}

\begin{abstract}

We study the ferminoic spectrum with Weyl correction,
which exhibits the non-Fermi liquid behavior.
Also, we find that both the height of the peak of the fermionic spectrum and the dispersion relation exhibit a nonlinearity with the variety of the Weyl coupling parameter $\gamma$,
which mean that such nonlinearity maybe ascribe to the one of the Maxwell field.
Another important property of this system is that for the holographic fermionic system with $\gamma<0$, the degree of the deviation from Fermi liquid is heavier
than that for the one with $\gamma>0$.
It indicates that there is a transition of coupling strength in the dual boundary field theory.

\end{abstract}

\maketitle

\section{Introduction}

The non-Fermi liquid phase usually involves strongly interactions,
in which the traditional perturbative tool loses its power.
We need resort to the alternative method to attack these problems.
AdS/CFT correspondence \cite{Maldacena:1997re,Gubser:1998bc,Witten:1998qj,Aharony:1999ti} provides such method.
By adding probe fermions over the Reissner-Nordstr$\ddot{\texttt{o}}$m (RN-AdS) back brane,
the dual non-Fermi liquid behavior, which exhibits non-linear dispersion relation, is observed in holographic framework \cite{Liu:2009dm}.
In particular, such scaling behavior is controlled by conformal dimensions in the IR CFT dual to the near horizon AdS$_2$ geometry \cite{Faulkner:2009wj}.
Subsequently, many extensive works on the holographic fermionic spectrum, especially the Fermi surface structure and associated excitations, have also been studied in
more general charged black brane geometries (see \cite{Cubrovic:2009ye,Wu:2011bx,Liu:2012tr,Ling:2013aya,Wu:2011cy,Gursoy:2011gz,Alishahiha:2012nm,
Fang:2012pw,Li:2012uua,Wang:2013tv,Wu:2013xta,Fang:2013ixa,Kuang:2014pna,Fang:2014jka,Fang:2015vpa,Fang:2015dia,Wu:2016hry,Alsup:2016fii} and references therein).
They provide more richer Fermi surface structure and can be the candidates for generalized non-Fermi liquids.

In this letter, we study the properties of the fermionic spectrum from a generalized Maxwell theory involving
the coupling between the Maxwell field strength and the Weyl tensor, which is a $4$ derivative theory.
The conductivity of this holographic system at zero charge density, which dual to the probe Maxwell theory in the Schwarzschild-AdS (SS-AdS) black brane geometry,
has been fully studied in
\cite{Myers:2010pk,Sachdev:2011wg,Hartnoll:2016apf,Ritz:2008kh,WitczakKrempa:2012gn,WitczakKrempa:2013ht,Witczak-Krempa:2013nua,Katz:2014rla,Bai:2013tfa}.
A peak or a dip in the low frequency optical conductivity can be observed depending on the sign of the parameter of the Weyl corrected term \cite{Myers:2010pk}.
This phenomena resembles the one in the superfluid-insulator quantum critical point (QCP) \cite{Myers:2010pk,Sachdev:2011wg,Hartnoll:2016apf}.
The transports in the higher derivative (HD) theory in the SS-AdS black brane geometry, in which the Maxwell field couples more Weyl tensors,
has also been explored in \cite{Witczak-Krempa:2013aea}.
The behavior of this system is very similar to that of the $O(N)$ $NL\sigma M$ in the large-$N$ limit \cite{Damle:1997rxu}.
Therefore, the HD theory in SS-AdS black brane geometry provides a possible road toward the QC dynamics described by certain CFTs.
The inhomogeneous disorder effect implemented by spatial linear dependent axionic fields has also been introduced
and the corresponding transport properties are also studied in \cite{Wu:2016jjd,Fu:2017oqa}.
The holographic superconductivity has also been explored in this HD framework (see \cite{Wu:2010vr,Ma:2011zze,Momeni:2011ca,Momeni:2012ab,Momeni:2012uc,Roychowdhury:2012hp,Zhao:2012kp,Momeni:2013fma,Momeni:2014efa,Zhang:2015eea,Mansoori:2016zbp,Ling:2016lis,Wu:2017xki} and references therein).
This model exhibits an interesting properties of
the running of superconducting energy gap approximately ranging from $5.5$ to $16$ \cite{Wu:2010vr,Wu:2017xki}.

Since we want to study the fermionic spectrum at finite charged density,
so we shall work over the charged AdS black brane geometry with Weyl corrections.
But it is hard to solve analytically or even numerically the equations when involving the backreaction from HD terms.
Alternatively, we can obtain the perturbative charged AdS black brane solution from Weyl corrections up to the first order of the coupling parameter
and the transports, diffusion and chaos, thermalization and entanglement entropy can be also explored at finite charge density \cite{Ling:2016dck,Li:2017nxh,Dey:2015poa,Dey:2015ytd,Mahapatra:2016dae}.
We shall explore the properties of the fermionic spectrum over this first order perturbative charged AdS black brane solution.

\section{The charged black brane with Weyl correction}

In this section, we shall give a brief review on the charged black hole with Weyl correction.
For the detail, please see \cite{Ling:2016dck}.
We start with the following action
\begin{eqnarray}
\label{A}
S=\int d^4x\sqrt{-g}\left[R+\frac{6}{L^2}-\frac{1}{4}F_{\mu\nu}F^{\mu\nu}+\gamma L^2C_{\mu\nu\rho\sigma}F^{\mu\nu}F^{\rho\sigma}\right]
\,,
\end{eqnarray}
where $F=dA$ is the Maxwell field strength and $C_{\mu\nu\rho\sigma}$ is the Weyl tensor.
$\gamma$ is the coupling parameter and is constrained in the region $\gamma\in[-\frac{1}{12},\frac{1}{12}]$ on the SS-AdS black brane background \cite{Myers:2010pk,Ritz:2008kh}.
Note that the factor of $L^2$ in action (\ref{A}) is introduced so that the coupling $\gamma$ is dimensionless.
But in what follows, we set $L=1$ for convenience.

Up to the first order of $\gamma$,
the Weyl corrected black brane solution is \cite{Ling:2016dck}
\begin{subequations}
\label{bbs}
\begin{align}
&
\label{metric}
ds^2=\frac{1}{z^2}(-f(z)dt^2+\frac{1}{f(z)}dz^2+g(z)(dx^2+dy^2))\,,
~~~~~
A=A_t(z)dt\,,
\
\\
&
f(z)=(1-z)p(z)\,,\ \ \ g(z)=1+\gamma\frac{\mu^2 z^4}{9}\,,
\
\\
&
p(z)=\gamma \frac{\mu^2 z^3}{180}
[240+2\mu^2(13z^4-14)+(\mu^2-100)(z^3+z^2+z)]
+1+z+z^2-\frac{\mu^2 }{4}z^3\,,
\
\\
&
A_t(z)=\mu\Big[(1-z)+\gamma\frac{z}{90}(180-29\mu^2+74\mu^2z^4-45z^3(4+\mu^2))\Big]\,,
\end{align}
\end{subequations}
where $\mu$ is the chemical potential in the dual boundary field theory.
The dimensionless Hawking temperature $\hat{T}\equiv T/\mu$, which is given by
\fa
\hat{T}=\frac{12-\mu^2}{16\pi\mu}+\gamma\frac{\mu(\mu^2-60)}{720\pi}\,.
\label{temh}
\ffa

Before proceeding, we shall analyze IR geometry of this Weyl corrected black brane,
which controls the low frequency behavior of holographic fermionic spectrum.
Here we only focus on the zero temperature limit, which is obtained by
\begin{eqnarray}
\label{mu}
\mu=\sqrt{\frac{3}{2}} \sqrt{-\frac{\sqrt{5} \sqrt{80 \gamma ^2+72 \gamma +45}}{\gamma
   }+\frac{15}{\gamma }+20}\,.
\end{eqnarray}
It is obvious that when $\gamma\rightarrow 0$, $\mu=2\sqrt{3}$, which recovers the case of RN-AdS black hole.
At extremal limit, this geometry flows to AdS$_2$ fixed point at IR with radius $L_2$ as
\fa
\label{L22}
L_2^2=
\frac{-4000 \gamma ^2+5 \sqrt{5} (40 \gamma +33) \sqrt{80 \gamma ^2+72 \gamma +45}-4956
   \gamma -2475}{4 \gamma }\,,
\ffa
which explicitly dependens on the Weyl parameter $\gamma$.

\section{Dirac equation}\label{DiE}

To study the fermionic spectrum, we use the following fermion action
\begin{eqnarray}
\label{actionspinor} S_{D}=i\int d^{4}x
\sqrt{-g}\,\overline{\zeta}\left(\Gamma^{a}\mathcal{D}_{a} - m \right)\zeta,
\end{eqnarray}
where $\mathcal{D}_{a}=\partial_{a}+\frac{1}{4}(\omega_{\mu\nu})_{a}\Gamma^{\mu\nu}-iq A_{a}$ is the covariant derivative and
$\slashed{F}=\frac{1}{2}\Gamma^{\mu\nu}(e_\mu)^a(e_\nu)^bF_{ab}$.
$(e_{\mu})^{a}$ is a set of
orthogonal normal vector bases and $(\omega_{\mu\nu})_{a}$ the spin connection.

From the action (\ref{actionspinor}),
the Dirac equation can be obtained as
\begin{eqnarray} \label{DiracEF}
\left[(\partial_{z}-m\sqrt{g_{zz}}\sigma^3)
+\sqrt{\frac{g_{zz}}{g_{tt}}}(\omega+qA_{t})i\sigma^2
+(-1)^{I} k \sqrt{\frac{g_{zz}}{g_{xx}}}
\right] F_{I} =0~,
\end{eqnarray}
with $I=1,2$.
We have made a change of $\zeta=(g_{tt}g_{xx}g_{yy})^{-\frac{1}{4}}\mathcal{F}$ and the Fourier expansion $\mathcal{F}\sim F(u,k)e^{-i\omega t + ikx}$.
\footnote{Due to the rotational symmetry in the spatial direction, we have set
$k_x=k$ and $k_y=0$.}
In addition,
we have used the following basis in the above equations
\begin{align}
\label{GammaMatrices}
 \Gamma^{z} = \left( \begin{array}{cc}
-\sigma^3 & 0  \\
0 & -\sigma^3
\end{array} \right), \;\;
 \Gamma^{t} = \left( \begin{array}{cc}
 i \sigma^1 & 0  \\
0 & i \sigma^1
\end{array} \right),  \;\;
\Gamma^{x} = \left( \begin{array}{cc}
-\sigma^2 & 0  \\
0 & \sigma^2
\end{array} \right).\;\;
\end{align}
Furthermore, by the decomposition
$
F_{I} \equiv (\mathcal{A}_{I}, \mathcal{B}_{I})^{T}
$, the above Dirac equations become
\begin{subequations}
\label{DiracEAB1}
\begin{align}
& (\partial_{z}-m\sqrt{g_{zz}})\mathcal{A}_{I}
+\sqrt{\frac{g_{zz}}{g_{tt}}}(\omega+qA_{t})\mathcal{B}_{I}
+(-1)^{I} k \sqrt{\frac{g_{zz}}{g_{xx}}}
\mathcal{B}_{I} =0~,
\
\\
& \label{DiracEAB2} (\partial_{z}+m\sqrt{g_{zz}})\mathcal{B}_{I}
-\sqrt{\frac{g_{zz}}{g_{tt}}}(\omega+qA_{t})\mathcal{A}_{I}
+(-1)^{I} k \sqrt{\frac{g_{zz}}{g_{xx}}}
\mathcal{A}_{I} =0~.
\end{align}
\end{subequations}
We package the above equations into flow equations, which is more convenient to implement the numerical computation,
\begin{eqnarray} \label{DiracEF1}
(\partial_{z}-2m\sqrt{g_{zz}}) \xi_{I}
+\left[ v + (-1)^{I} k \sqrt{\frac{g_{zz}}{g_{xx}}}  \right]
+ \left[ v - (-1)^{I} k \sqrt{\frac{g_{zz}}{g_{xx}}}  \right]\xi_{I}^{2}
=0
~,
\end{eqnarray}
where $\xi_{I}\equiv \frac{\mathcal{A}_{I}}{\mathcal{B}_{I}}$
and $v=\sqrt{\frac{g_{uu}}{g_{tt}}}(\omega+q
A_{t})$.
To solve the above flow equation, we shall impose the ingoing boundary condition at the horizon $z=1$, which gives $\xi_{I}=i$ for $\omega\neq0$.
For the case of $\omega=0$, please see \cite{Liu:2009dm}.
And then, the boundary Green's function can be read off as \cite{Liu:2009dm}
\begin{eqnarray} \label{Green}
G (\omega,k)= \lim_{u\rightarrow 0} z^{-2m}
\left( \begin{array}{cc}
\xi_{1}   & 0  \\
0  & \xi_{2} \end{array} \right)  \ .
\end{eqnarray}

Next, we explore the characteristics of fermionic spectrum with Weyl correction by numerically solving Eq. (\ref{DiracEF1}).
Here the main concern will focus on how does the Weyl parameter $\gamma$ affect the characteristics of the fermionic spectrum.

\section{Holographic fermionic spectrum}

\begin{figure}
\center{\includegraphics[scale=0.26]{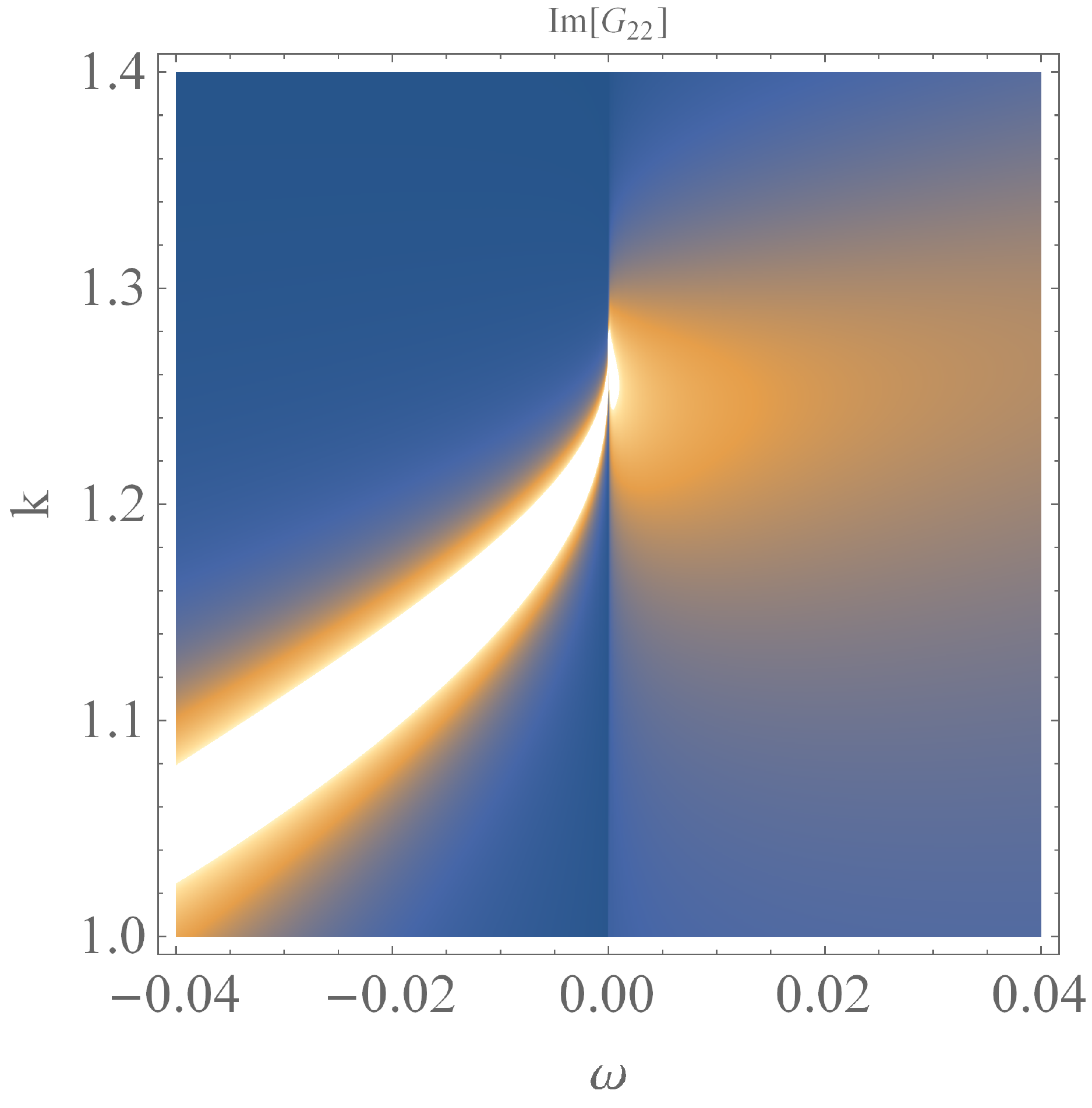}\ \hspace{0.2cm}
\includegraphics[scale=0.26]{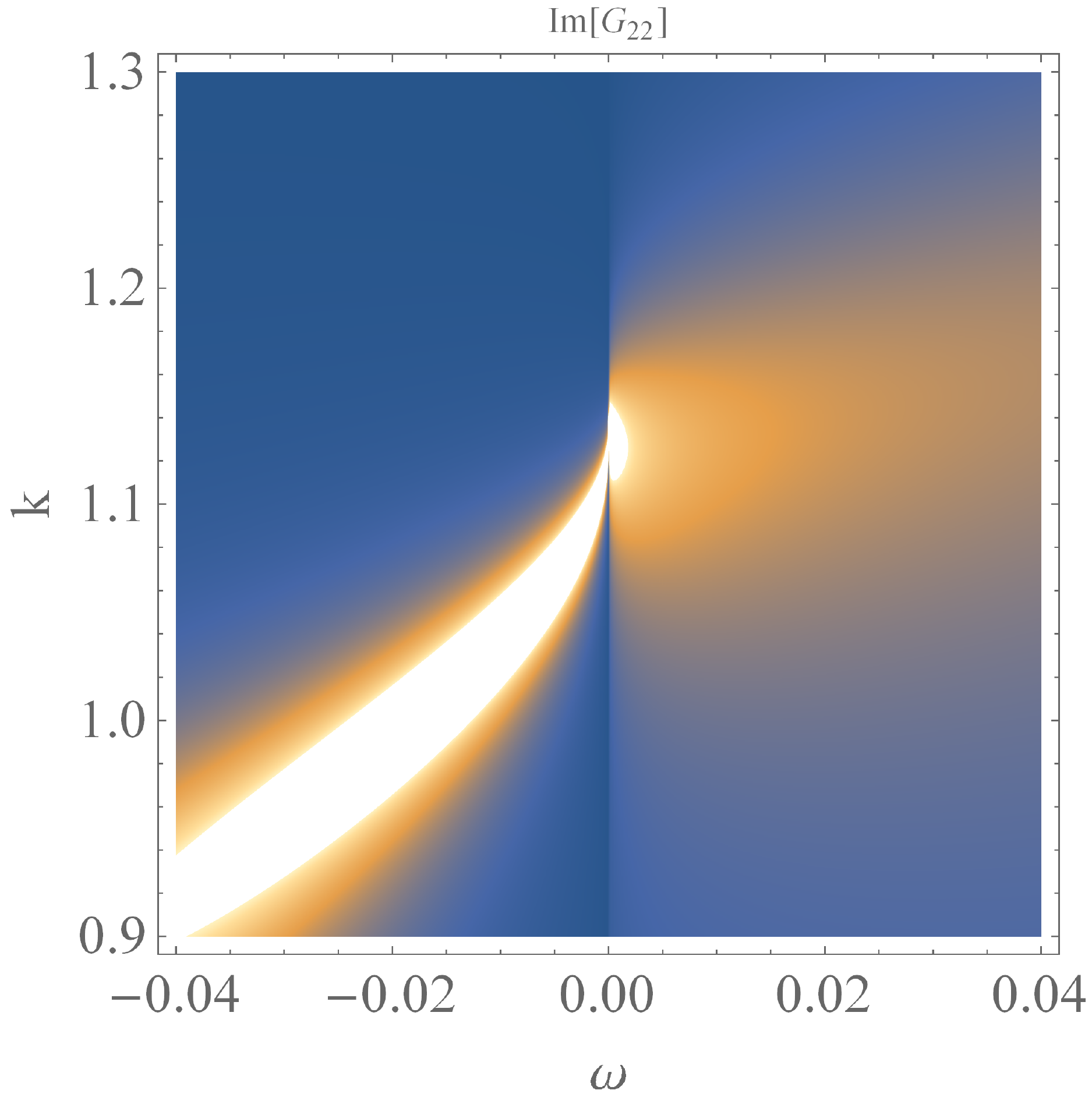}\ \\
\includegraphics[scale=0.26]{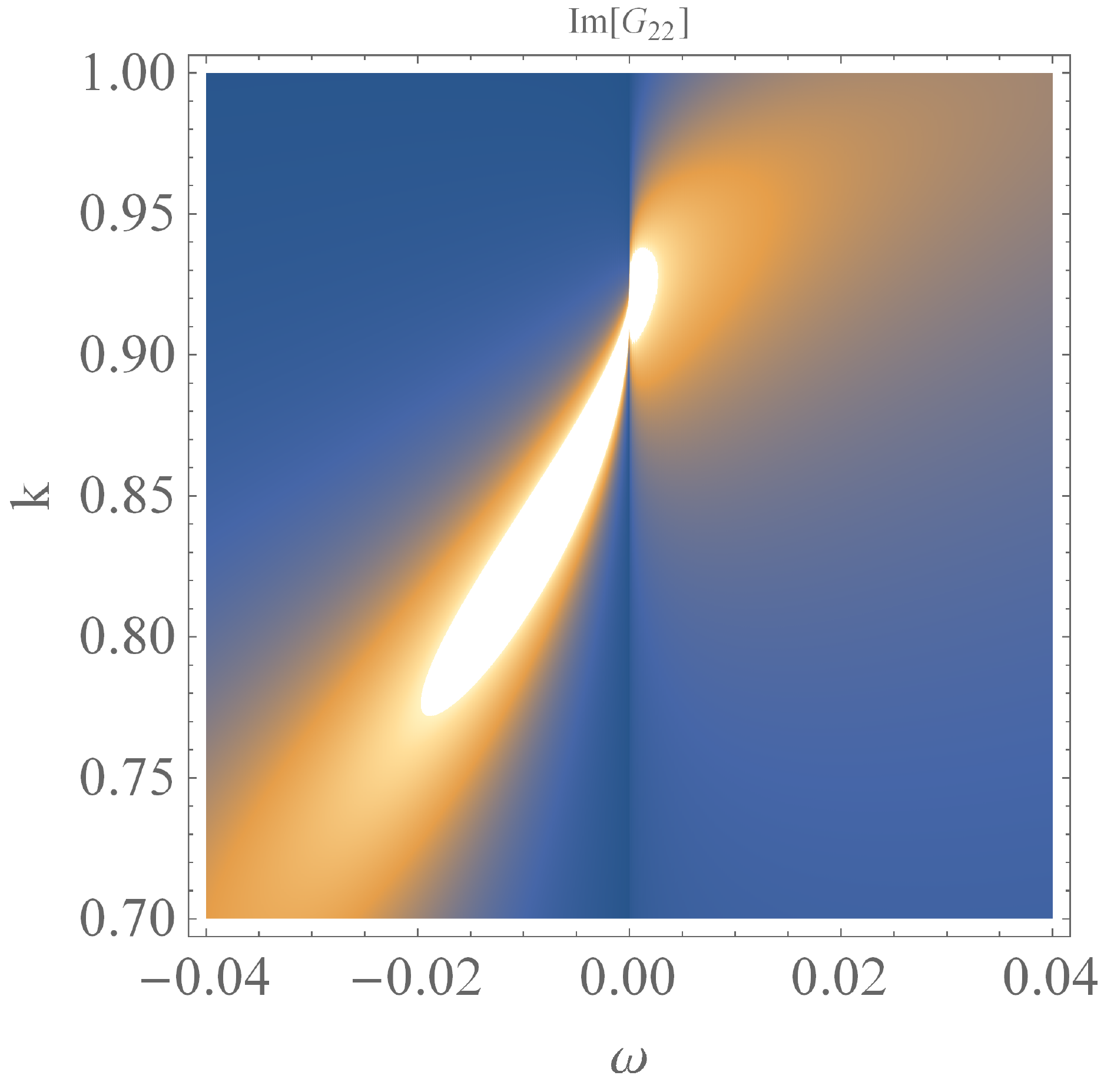}\ \hspace{0.2cm}
\includegraphics[scale=0.26]{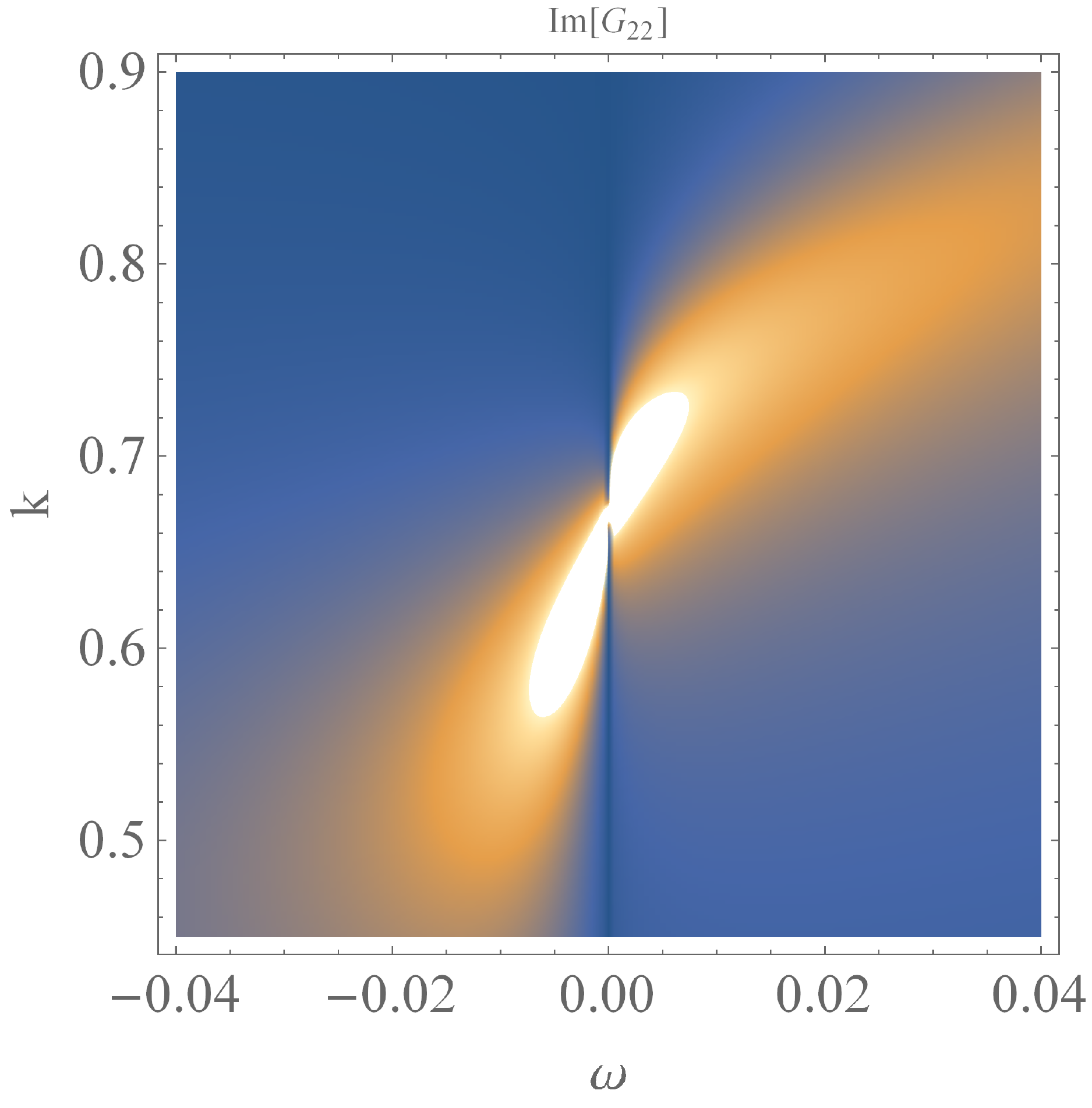}\ \hspace{0.2cm}
\includegraphics[scale=0.26]{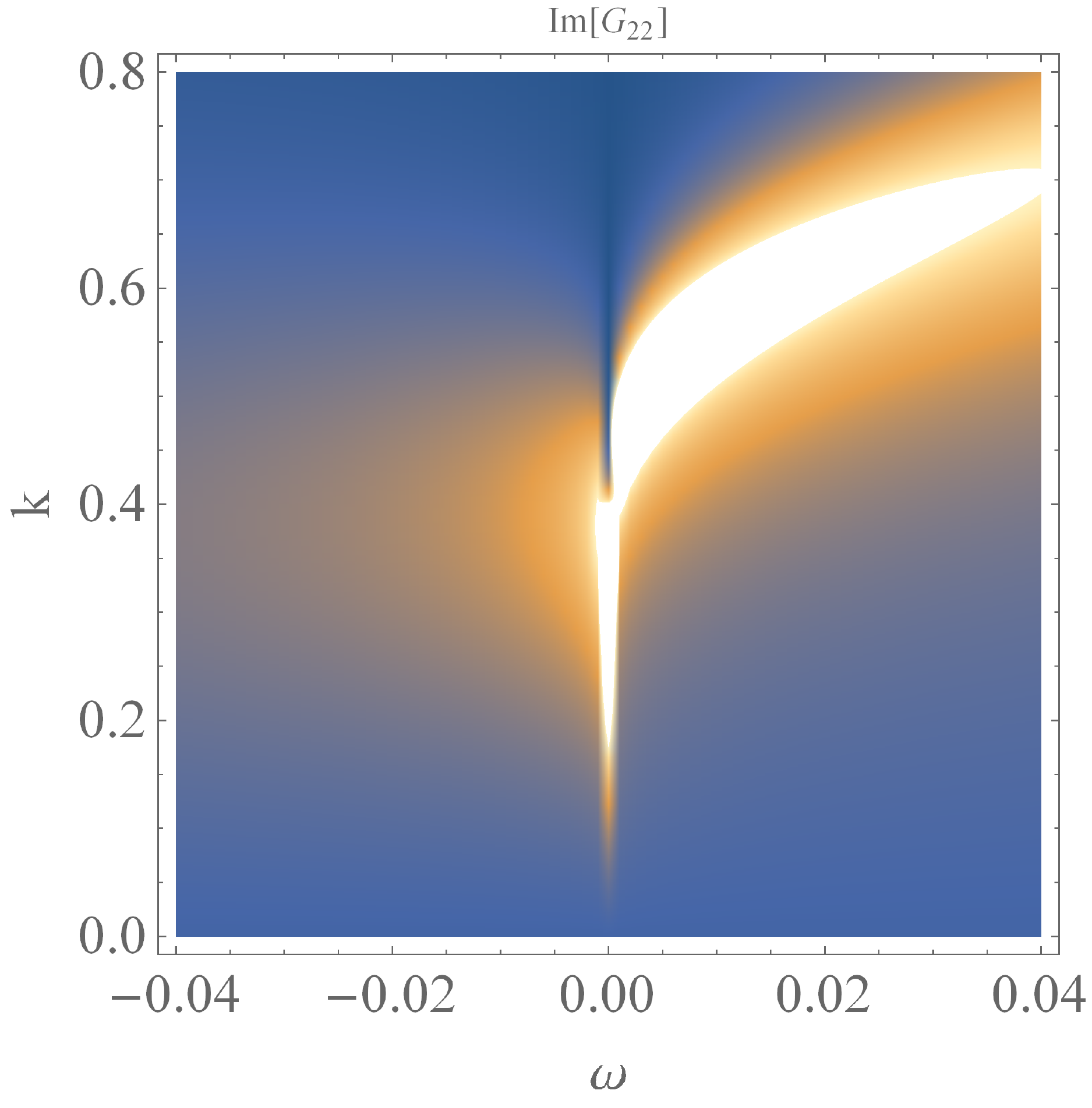}\ \\
\caption{\label{DP}The density plots of ImG$_{22}(\omega,k)$ for $\alpha=-\frac{1}{12}$ (the left plot above),
$\alpha=-0.05$ (the right plot above), $\alpha=0$ (the first plot below), $\alpha=0.05$ (the second plot below) and $\alpha=\frac{1}{12}$.
All the other parameters are fixed as $q=\frac{1}{2}$, $m=0$ and $T=0$.}}
\end{figure}

\begin{figure}
\center{
\includegraphics[scale=0.5]{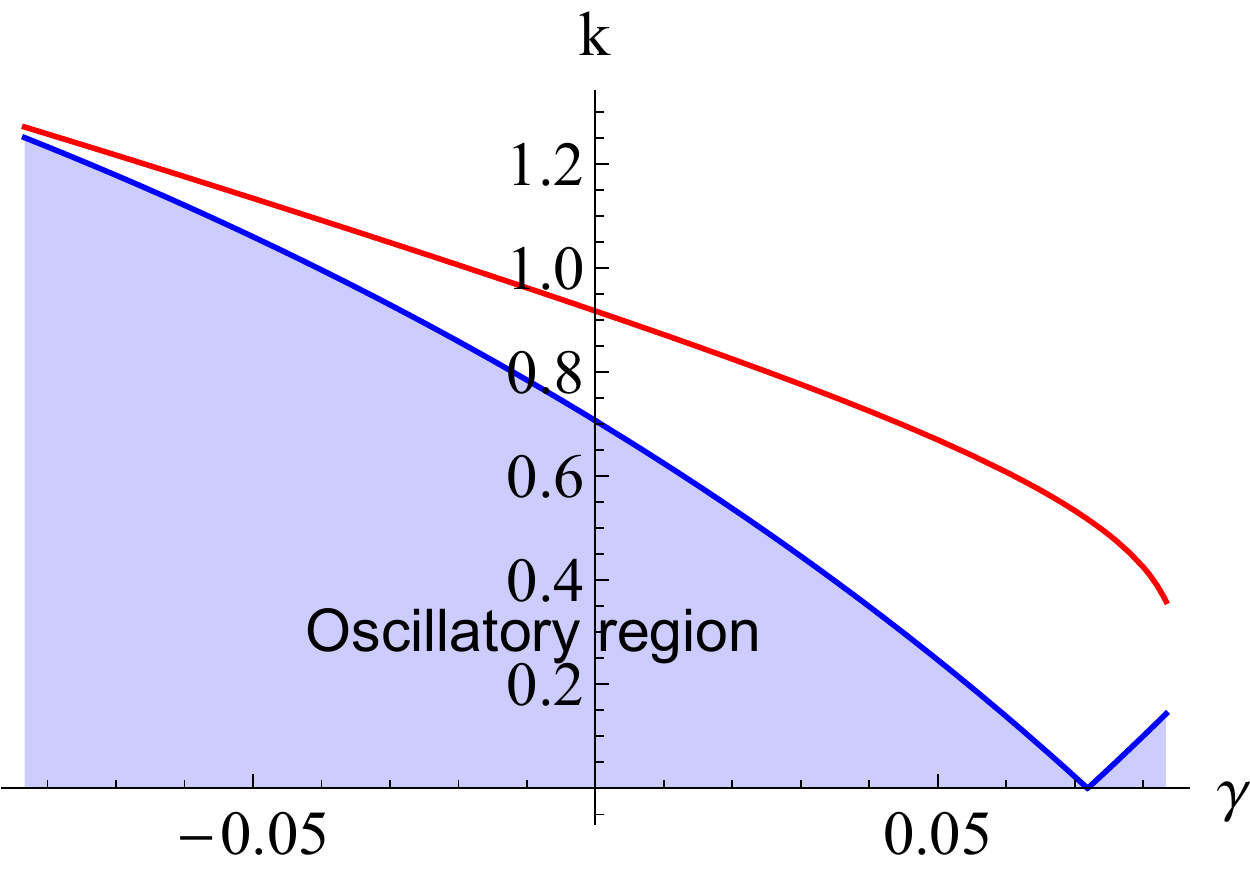}\ \hspace{0.8cm}
\includegraphics[scale=0.7]{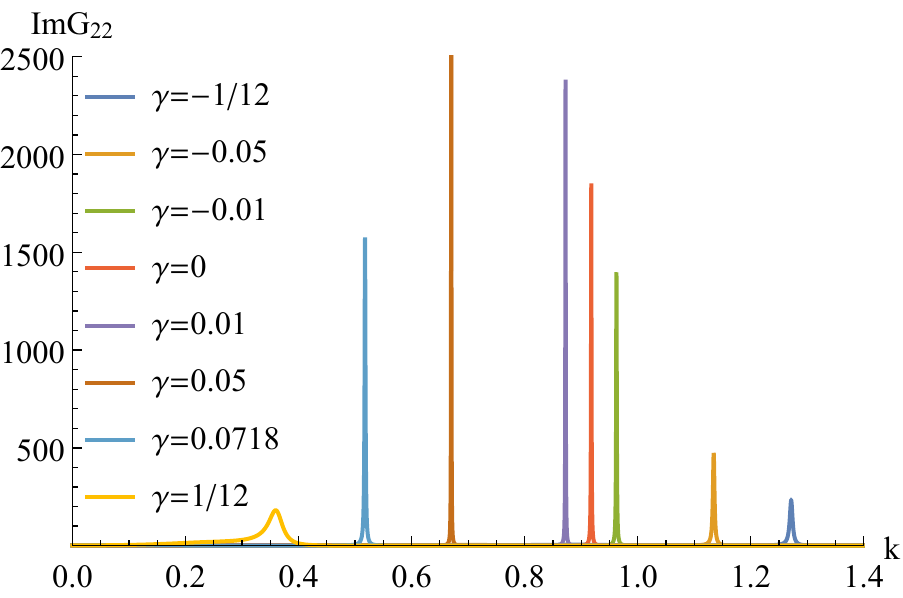}\ \\
\caption{\label{kFvsalphs}Left plot: The Fermi momentum $k_F$ vs. $\gamma$.
The blue zone is the oscillatory region.
Right plot: ImG$_{22}$ as the function of $k$ at $\omega=0$ with different Weyl parameters $\gamma$.
}}
\end{figure}

\begin{figure}
\center{
\includegraphics[scale=0.5]{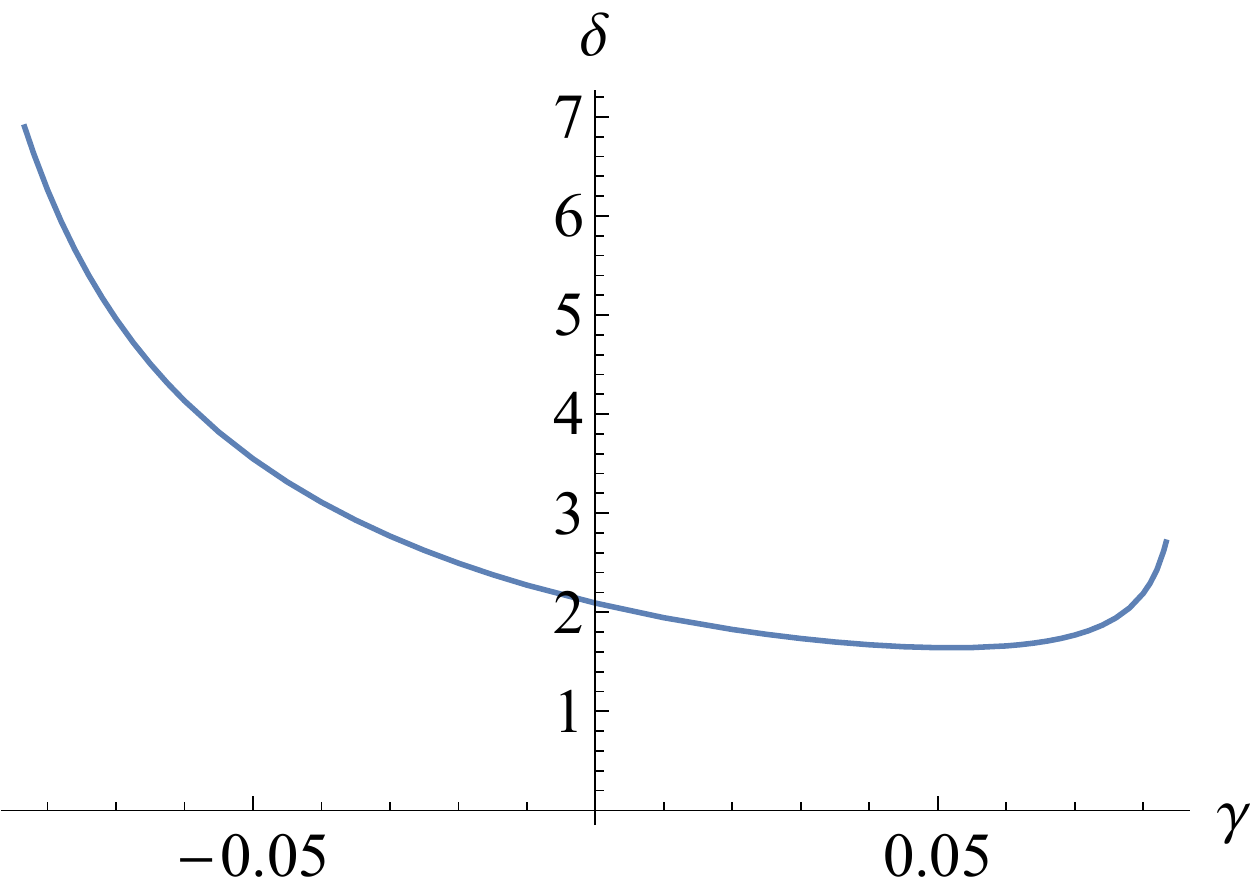}\ \hspace{0.8cm}
\includegraphics[scale=0.78]{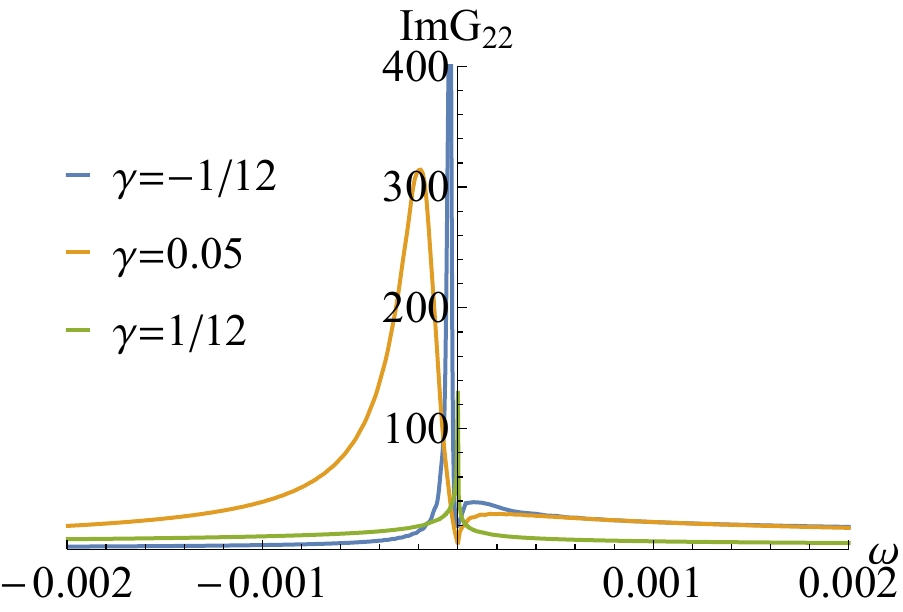}\ \\
\caption{\label{disvsa}Left plot: The scaling exponent $\delta$ of the dispersion relation as the function of Weyl parameter $\gamma$.
Right plot: ImG$_{22}$ as the function of $\omega$ for fixed $k$ ($\hat{k}<0$) with different Weyl parameters $\gamma$.
}}
\end{figure}

Without loss of generality, we set $q=\frac{1}{2}$, $m=0$, $T=0$ so that we can only concentrate on the effect of Weyl correction on the fermionic spectrum.
Firstly, we show the density plots of ImG$_{22}$ in FIG.\ref{DP},
from which we can see that a quasi-particle-like peak near $\omega=0$ emerges in the region of Weyl parameter $\gamma$ allowed.
Usually, its location in momentum space corresponds to Fermi momentum $k_F$ ($\omega=0$).
Quantitatively, we can numerically work out the Fermi momentum $k_F$ by locating the peak in momentum space at $\omega=0$.
The results are summarized in Table \ref{TkFdelta} and FIG.\ref{kFvsalphs},
from which we see that $k_F$ decreases monotonously with the increase of $\gamma$
\footnote{The blue zone in left plot in FIG.\ref{kFvsalphs} is the oscillatory region,
in which $\nu_I(k)$ is pure imaginary \cite{Liu:2009dm,Faulkner:2009wj}.
We note that at given region of parameters, the Fermi momentum is outside the oscillatory region
so that all peak possess the meaning of Fermi surface.}.
Furthermore, we show ImG$_{22}$ as a function of $k$ at $\omega= 0$ for sample Weyl parameter $\gamma$ (right plot in FIG.\ref{kFvsalphs}).
It exhibits a nonlinearity of the height of peak with the variety of $\gamma$, which is similar with that found in fermionic spectrum in BI-AdS black hole \cite{Wu:2016hry}
but different from that found in other higher curvature correction to Einstein gravity, for instance, the Gauss-Bonnet case \cite{Wu:2011bx}.
As have pointed out in \cite{Wu:2016hry}, such nonlinearity may be ascribe to the one of the Maxwell field.
This nonlinearity also reflects in the dispersion relation, which is non-monotonous change with $\gamma$ (Table \ref{TkFdelta} and the left plot in FIG.\ref{disvsa}).
In addition, we also note that the dispersion relation is nonlinearity for the arbitrary $\gamma$,
which indicates that our fermionic system with Weyl correction is non-Fermi liquid system.

\begin{widetext}
\begin{table}[ht]
\begin{center}
\begin{tabular}{|c|c|c|c|c|c|c|c|c|c|}
         \hline
~$\gamma$~ &~$-\frac{1}{12}$~&~$-0.05$~&~$-0.01$~&~$0$~&~$0.01$~&~$0.05$~&~$\frac{1}{12}$~
          \\
        \hline
~$k_F$~ &~$1.2716$~&~$1.1345$~&~$0.9624$~&~$0.9178$~&~$0.8723$~&~$0.6701$~&~$0.3590$~
          \\
        \hline
~$\delta$~ & ~$6.8986$~&~$3.5499$~&~$2.2739$~ & ~$2.0926$~&~$1.9449$~&~$1.6421$~&~$2.7119$~
          \\
        \hline
\end{tabular}
\caption{\label{TkFdelta}The Fermi momentum $k_F$ and scaling exponent $\delta$ of dispersion relation
with different Weyl parameter $\gamma$.}
\end{center}
\end{table}
\end{widetext}


We are interested in the behavior of ImG$_{22}$ in the region of small $\hat{k}\equiv k-k_F$ and $\omega$.
The right plot in FIG.\ref{disvsa} shows ImG$_{22}$ as the function of $\omega$ for fixed $k$ ($\hat{k}<0$) with different Weyl parameters $\gamma$.
We can observe that with the exception of the highest bound of $\gamma$ ($\gamma=\frac{1}{12}$), for $\hat{k}<0$,
a sharp quasi-particle-like peak appears in the region $\omega<0$
and a small bump in $\omega>0$ (also see FIG.\ref{DP}).
It is the same as the case in RN-AdS background \cite{Liu:2009dm} and implies
that a quasi-particle-like pole appears in the left quadrant of the lower-half complex $\omega$-plane.
For $\gamma=\frac{1}{12}$, the quasi-particle-like peak locates in $\omega=0$ (FIG.\ref{DP} and the right plot in FIG.\ref{disvsa}),
which is very different from that found in RN-AdS background \cite{Liu:2009dm}
or other geometries \cite{Wu:2011bx,Wu:2011cy,Li:2012uua,Wu:2013xta,Kuang:2014pna}.
It deserves to further explore the reason behind this phenomenon.

We also note that for the holographic fermionic system with $\gamma<0$, the degree of the deviation from Fermi liquid is heavier
than that for the one with $\gamma>0$, which means that the coupling strength of the dual boundary field theory with $\gamma<0$ is stronger
than that with $\gamma>0$.
This result is consistent with that observed in \cite{Myers:2010pk} there is a transition from a Drude-like peak for $\gamma>0$ to a dip for $\gamma<0$
and that in \cite{Wu:2010vr,Wu:2017xki,Ma:2011zze,Mansoori:2016zbp} there is a running of superconductivity energy gap varying from $5.5$ to $16$,
which are also the manifestation of the transition of coupling strength.

\section{Conclusion and discussion}\label{sec-conclusion}

In this paper, we study the ferminoic spectrum with Weyl correction,
which exhibits the non-Fermi liquid behavior.
In addition, by studying the imaginary part of the Green function as the function of $k$ at $\omega=0$,
we find that the height of peak exhibits a nonlinearity with the variety of the Weyl coupling parameter $\gamma$.
Such nonlinearity is also observed in the dispersion relation.
The observations are similar with that found fermionic spectrum in BI-AdS black hole \cite{Wu:2016hry}
and further confirm that such nonlinearity exhibited in the fermionic spectrum originates from the one of the Maxwell field.

Another important property of this system is that for the holographic fermionic system with $\gamma<0$, the degree of the deviation from Fermi liquid is heavier
than that for the one with $\gamma>0$.
It indicates that there is a transition of coupling strength in the dual boundary field theory.
This observation is consistent with that in the conductivity of the boundary field theory dual to the Maxwell theory with Weyl correction in SS-AdS black brane \cite{Myers:2010pk}
and that in the running of superconductivity energy gap \cite{Wu:2010vr,Wu:2017xki,Ma:2011zze,Mansoori:2016zbp}.

In future, we can add the dipole coupling term in the fermionic action and study its fermionic response
to see the effects of the Weyl coupling on the formation of Mott gap.
Also, we can explore the non-relativistic fermionic system with Weyl correction.
The related works are under progress.

\section*{Acknowledgements}

This work is supported by the Natural Science Foundation of China under
Grant Nos. 11775036 and 11305018, and by Natural Science Foundation of Liaoning Province under
Grant No.201602013.

\end{document}